\documentclass[reprint, superscriptaddress, amsmath, amssymb, aps, prb,nofootinbib]{revtex4-2}

\usepackage[utf8]{inputenc}
\usepackage{graphicx}
\usepackage{hyperref}
\usepackage{xcolor}
\usepackage{braket}
\usepackage{bm}

\newcommand{\UV}{\mathrm{UV}}
\newcommand{\ketbra}[2]{\ket {#1} \hskip -0.8ex \bra {#2}}
\newcommand{\expct}[1]{\left\langle #1 \right\rangle}
\newcommand{\MF}{\mathrm{MF}}

\DeclareMathOperator{\tr}{tr}

\begin{document}

\title{Thermal Multi-scale Entanglement Renormalization Ansatz for Variational Gibbs State Preparation}

\author{Troy J. Sewell}
\email{tjsewell@umd.edu}
\affiliation{Joint Center for Quantum Information and Computer Science, University of Maryland, College Park, Md, 20742}

\author{Christopher David White}
\affiliation{Joint Center for Quantum Information and Computer Science, University of Maryland, College Park, Md, 20742}

\author{Brian Swingle}
\affiliation{Department of Physics, Brandeis University, Waltham, Massachusetts, 02453}
    
\begin{abstract}
    Many simulation tasks require that one first prepare a system's Gibbs state.
    We present a family of quantum circuits for variational preparation of thermal Gibbs states on a quantum computer;
    we call them the \textit{thermal multi-scale entanglement renormalization ansatz} (TMERA).
    TMERA circuits transform input qubits to wavepacket modes localized to varying length scales and approximate a systems Gibbs state as a mixed state of these modes. 
    The TMERA is a based on the deep multi-scale entanglement renormalization ansatz (DMERA) 
    a TMERA modifies a ground-state DMERA circuit by preparing each input qubit as a mixed state. The excitation probabilities for input qubits serve as variational parameters used to target particular temperature Gibbs states.  
    Since a TMERA is a special case of the product spectrum ansatz for thermal states 
    it is simple to prepare, analyze, and optimize.
    We benchmark the TMERA on the transverse field Ising model in one dimension
    and find that for $D=6$ it produces global fidelities $\mathcal F > 0.4$ for 512-site systems across all temperatures.
\end{abstract}

\maketitle

\section{Introduction}

Quantum computers promise efficient simulation of quantum physics \cite{feynman_simulating_1982}.
Nearly any quantum simulation task requires the preparation of some physically motivated state to be measured or used as the initial state of a dynamical simulation.
States in thermal equilibrium, although dynamically trivial, are of general importance to many physical problems.

\begin{figure}[t]
  \centering
  \begin{minipage}{0.8\linewidth}
  \centering \includegraphics[width=\linewidth]{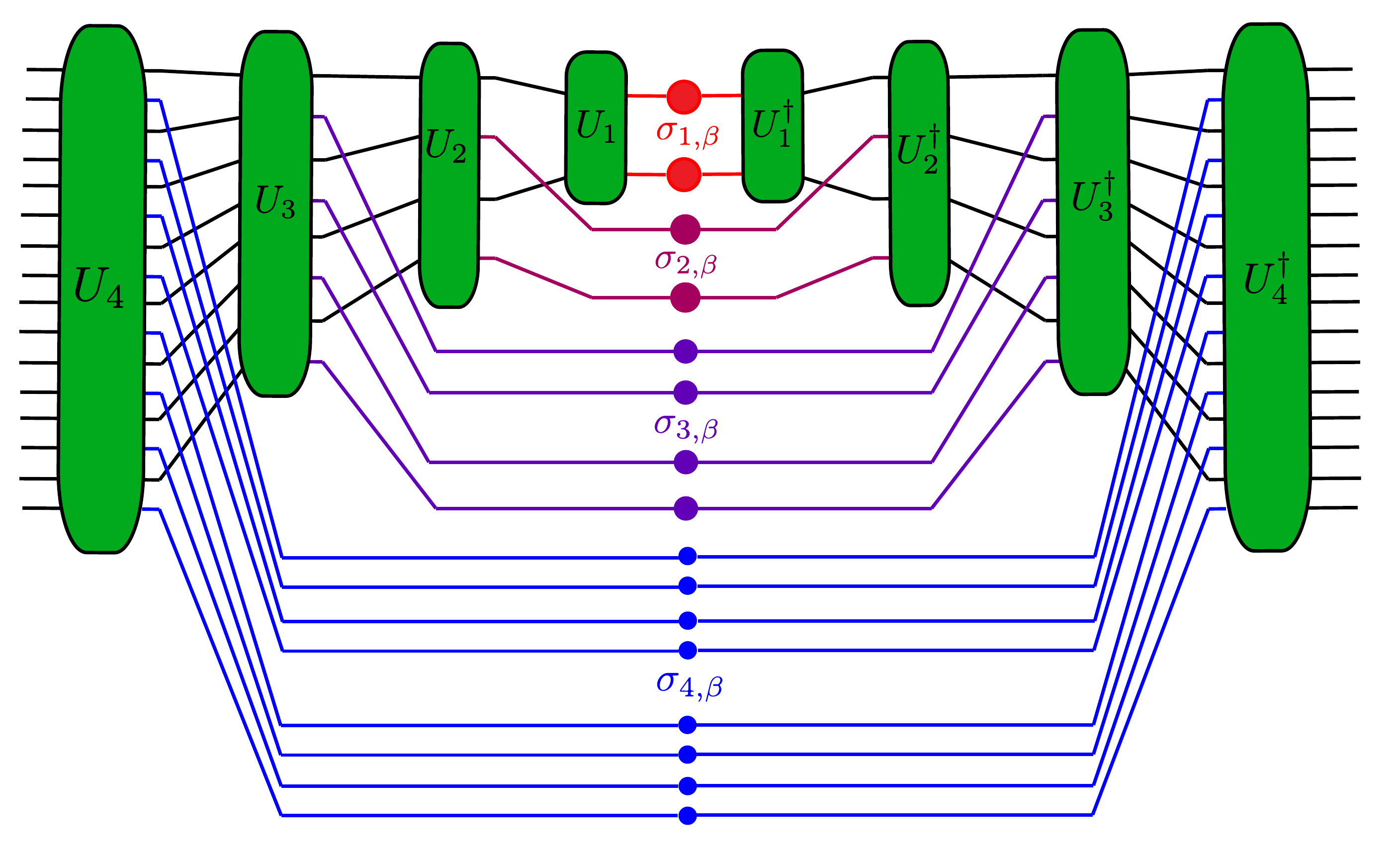}
 \end{minipage}\\
   \begin{minipage}{0.48\linewidth}
  \includegraphics[width=\linewidth]{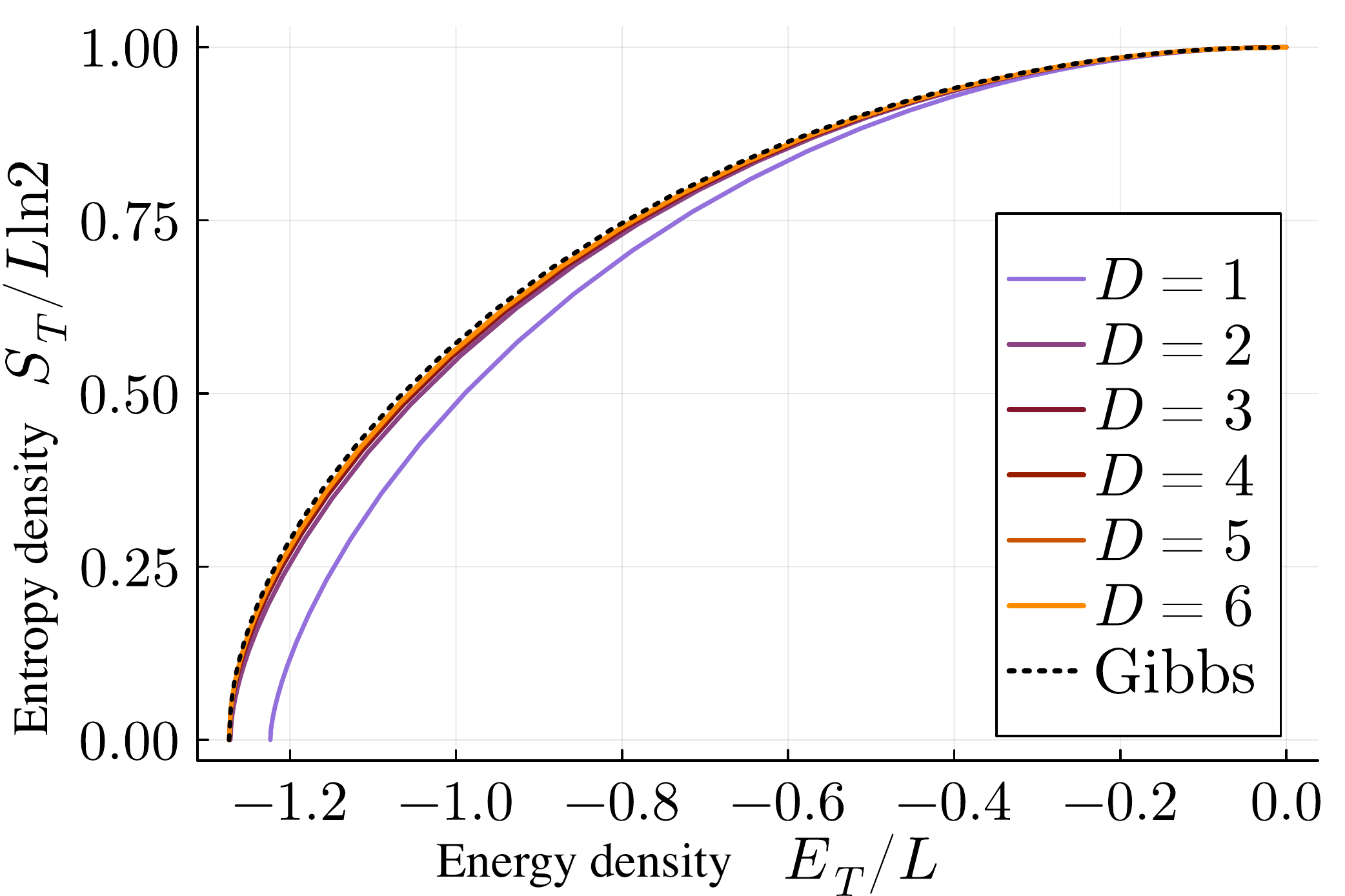}
 \end{minipage}\hfill
   \begin{minipage}{0.48\linewidth}
  \includegraphics[width=\linewidth]{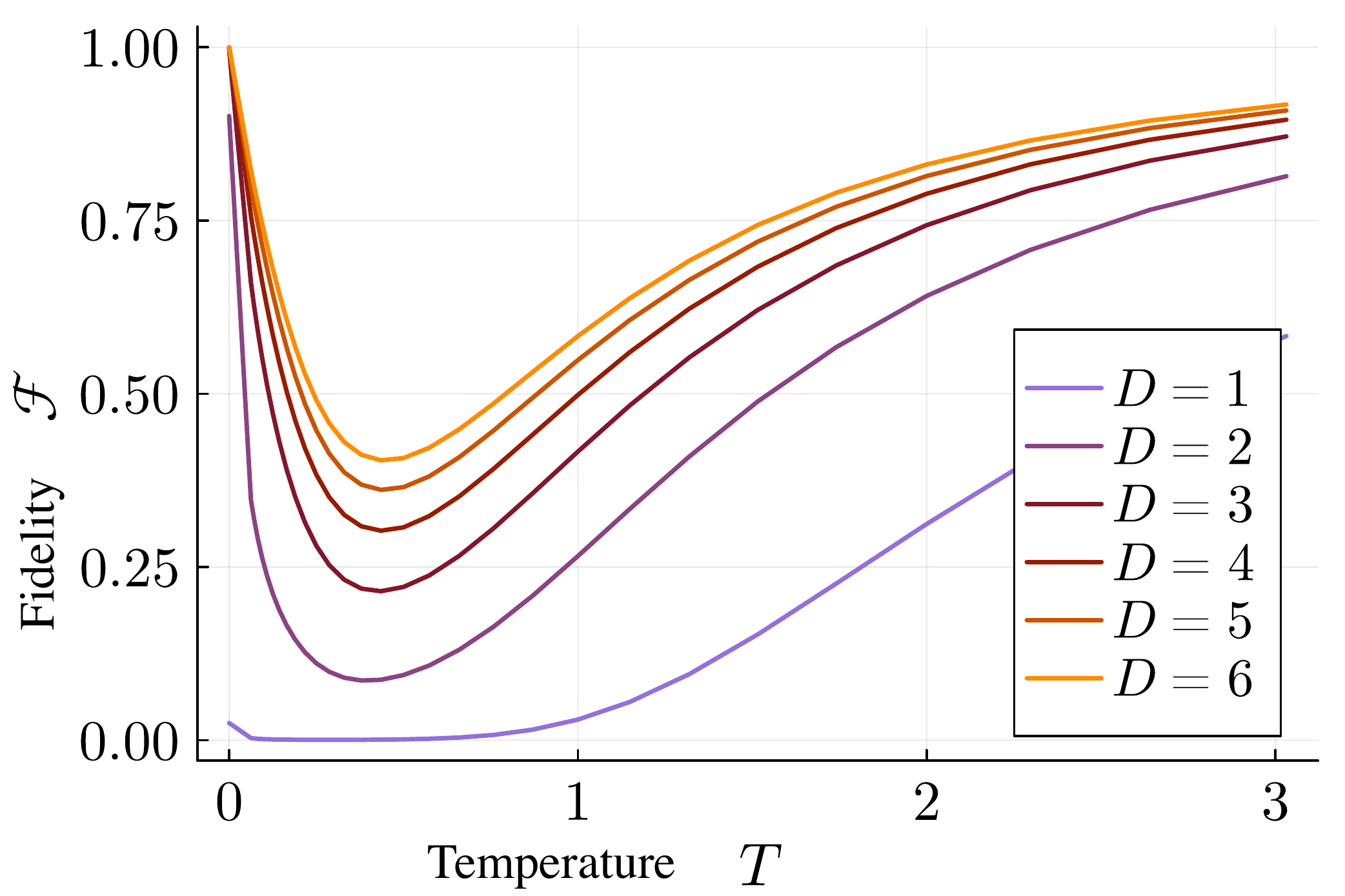}
 \end{minipage}
  \caption{\textbf{Top:} The TMERA circuit. A unitary DMERA quantum circuit optimized for ground state preparation (green) is applied to a product of mixed qubit states (colored dots).
  The circuit ansatz separates energies by scale, with low energy qubits introduced in very mixed/hot states (red) and high energy qubits introduced in relatively pure/cold states (blue). 
  \textbf{Bottom:} The global state energy and entropy (left) and state fidelity (right) for an $L=512$ spin system show that the ansatz is able to prepare accurate Gibbs state approximations for the critical Ising chain at a variety of temperatures.
  The ansatz improves with the circuit depth $D$ of the scale transformations.
  }
  \label{fig:highlights}
 \end{figure}

But preparing Gibbs states on a quantum computer is in general not easy.
The problem is NP complete, because computational problems can be encoded in statistical mechanics models,
and QMA complete because in the low-temperature limit the Gibbs state becomes the ground state \cite{kempe_complexity_2006,brown_computational_2011,bravyi_complexity_2021}.
Indeed any algorithm preparing Gibbs states \cite{terhal_problem_2000,poulin_sampling_2009,bilgin_preparing_2010,temme_quantum_2011,yung_quantumquantum_2012,riera_thermalization_2012,chowdhury_quantum_2016,kastoryano_quantum_2016,brandao_finite_2019,cottrell_how_2019,verdon_quantum_2019,Wu_2019,Martyn_2019,chowdhury_variational_2020,motta_determining_2020,zhu2020generation,bauer_quantum_2020,sagastizabal_variational_2020,liu_solving_2021,sun_quantum_2021,guo_thermal_2021,cerezo_variational_2021,wild_quantum_2021,wang_variational_2021,silva_fragmented_2022,davoudi_toward_2022}
is only efficient for a restrictive subclass of problems (e.g. Hamiltonians with local commuting terms  \cite{kastoryano_quantum_2016} or short-range correlated states \cite{brandao_finite_2019}),
and many face challenges like barren plateaus,
complicated energy or entropy measurement schemes,
or reliance on oracles that themselves require expensive implementations.

We seek to prepare Gibbs states on current and near-future quantum computers.
These computers will be NISQ (noisy intermediate-scale quantum \cite{preskill_quantum_2018}) devices,
limited in the number of qubits available and the fidelity with which quantum operations can be performed on those qubits.
One can avoid accumulation of errors on these devices by working within a family of short-depth circuits---an ansatz for the state to be prepared.
Parameters for a circuit preparing the target state
can then be found through a process of hybrid quantum-classical variational optimization,
in which quantum measurements are used by a classical optimization algorithm
to determine the optimal circuit parameters.
For example, to prepare the ground state of a Hamiltonian
one would measure the energy for different circuit parameters
in order to classically estimate energy gradients
and perform gradient descent on the circuit parameters.
For ground states this approach is known as a \textit{variational quantum eigensolver} (VQE)
and has been demonstrated for small systems on existing quantum computing devices \cite{Peruzzo_2014, zhu2020generation, Ho_2019}.
More generally, \textit{variational quantum algorithms} (VQA) can be used for other types of near-term applications, 
where a small ansatz circuit is optimized to prepare a desired state \cite{cerezo_variational_2021}.
For Gibbs states, the free energy can be used as the scalar quantity to be optimized\cite{Wu_2019}.
But measuring a state's free energy is difficult, because it requires measuring the state's entropy.

We suggest that for many practical applications,
the Gibbs state preparation problem becomes tractable
when we take advantage of \textit{scale separation}%
---the fact that in physical Hamiltonians length scales are associated with energy scales.
The separation of degrees of freedom at different length scales is made manifest in multi-scale quantum circuits.
This motivates our variational ansatz for Gibbs states, which we call the \textit{thermal multi-scale renormalization ansatz} (TMERA).

Multiscale circuits like MERAs or DMERAs progressively enlarge the Hilbert space by tensoring on new qubits in the $\ket 0$ state.
This progressive enlargement implements a fine-graining scaling transformation.
The TMERA instead progressively enlarges the Hilbert space by tensoring on new qubits in mixed states with entropy controlled by the scale at which the new qubits are introduced, seen at the top of Fig. \ref{fig:highlights}.
The mixed state qubits can be prepared by independently sampling computational basis states on each qubit, with a scale-dependent sampling bias that sets the entropy at a given energy range. Since each qubit is introduced at a given length scale, and therefore energy scale, a probability can be associated to it for a target temperature.
Short length scale (UV) degrees of freedom have high energy,
hence lower excitation probability,
while long length scale (IR) modes have low energy, hence higher excitation probability.
This is an example of a product spectrum ansatz for preparing Gibbs states \cite{Martyn_2019}, where multi-scale circuits are used to localize the input product state in energy. 

A product spectrum ansatz allows us to know the entropy of the final state a priori, since the unitary quantum circuit does not change the entropy of the initial product state. This avoids the pitfall of having to estimate the entropy via quantum measurements which can be practically prohibitive.  

Despite the limitations of using a low-depth variational circuit with a product state spectrum, we find that high fidelity approximations to Gibbs states can be found for large systems at various temperatures using few variational parameters.
We benchmark our ansatz with the transverse-field Ising model in one dimension at its critical point.
Scaling and translation symmetry of this model strongly constrains the system's Hamiltonian and correlations, and therefore can greatly simplify the underlying quantum circuit ansatz, requiring only a constant number of variational parameters the DMERA ground state circuit.
Scale-invariance of critical models also constrains the energy density to scale according to a fixed critical exponent. This reduces the ansatz for excitation probabilities to two variational parameters which set the critical exponent and the temperature. However, the energy associated to each scale of the DMERA may also be measured experimentally, which effectively solves the Gibbs state problem in the context of our ansatz by fixing a single parameter family of states which approximate Gibbs states of all temperatures. 
For the critical Ising model we find that the TMERA ansatz gives a global fidelity $> 0.5$ on 256 qubits using a log depth quantum circuit.

We structure the paper as follows.
In Sec.~\ref{s:ansatz-intuition} we describe our guiding intuition and present the TMERA circuit ansatz.
In Sec.~\ref{s:model} we give the model we use for benchmarking (the transverse field Ising model),
in Sec.~\ref{s:properties} we discuss properties of the TMERA, including the energy and entropy measurements required for variational optimization of the free energy,
and in Sec.~\ref{s:benchmarking} we benchmark our model by measuring its fidelity with the true thermal state,
its entropy as a function of energy, 
and other physical quantities.

\section{Background, intuition, and circuit ansatz}\label{s:ansatz-intuition}

\subsection{Background and intuition}

Before contemplating mixed state preparation, let us contemplate pure state preparation.
Suppose we wish to prepare the ground state of some Hamiltonian on a quantum computer,
without having a circuit in hand to do the preparation.
We can construct a family of parameterized quantum circuits which are within the capabilities of the available NISQ device and serve as an ansatz for the state we want to prepare.
The (approximate) ground state circuit is then the circuit within the ansatz family which prepares the state with lowest energy. We can find the desired circuit through a hybrid quantum-classical optimization process
in which the  quantum computer evaluates the energy at a given set of parameters and the classical computer uses those measurements to update the circuit parameters, through some optimization such as gradient descent, until the optimal circuit parameters are found. This is known as a variational quantum eigensolver (VQE) \cite{Peruzzo_2014, McClean_2016, VQE_review} when the objective function is a ground state energy, but the hybrid variational approach can be applied to other kinds of objective functions for more general quantum state preparation.
The quantum approximate optimization algorithm (QAOA) \cite{QAOA, QAOA2}, for instance,
has been applied to combinatorial optimization problems like MaxCut.

The art of a successful variational algorithm is finding a circuit ansatz that is expressive enough to represent a state close to the target state, but simple enough to be effectively optimized and implemented given the limitations of existing quantum technology.
Physical intuition can guide us in designing a circuit ansatz for physical problems%
---for example, a short-range correlated state is a good match for a low-depth brickwork circuit.
If the state is translation invariant, the circuit may likewise be constrained to be spatially homogeneous or periodic in order to reduce the ansatz complexity.

If the model is also scale invariant, perhaps because it sits at a quantum critical point,
we should construct an ansatz that is likewise scale-invariant. 
The \textit{multi-scale entanglement renormalization ansatz} (MERA) is such a scale-invariant ansatz,
initially developed to target critical ground states using tensor netowrks.\cite{Vidal_2007, EV09} 
The most basic MERA prepares the state in a scale-by-scale manner,
preparing a sequence of states $\{\ket{\psi_\ell}\}_\ell$, each on $2^\ell$ qubits.
The $\ket{\psi_\ell}$ can be viewed as a coarse-grained version of the target state.
The MERA, then, consists of a series of identical ``fine-graining'' layers, known as a \textit{scaling transformation}.
Each layer maps an incoming Hilbert space of $2^{\ell-1}$ qubits to an outgoing Hilbert space of $2^\ell$ qubits;
it does this by applying an isometry to each qubit that embeds that qubit's Hilbert space in a larger Hilbert space,
and then applying unitaries between neighboring qubits to tune local correlations.

\begin{figure}
    \centering
      \includegraphics[width=\linewidth]{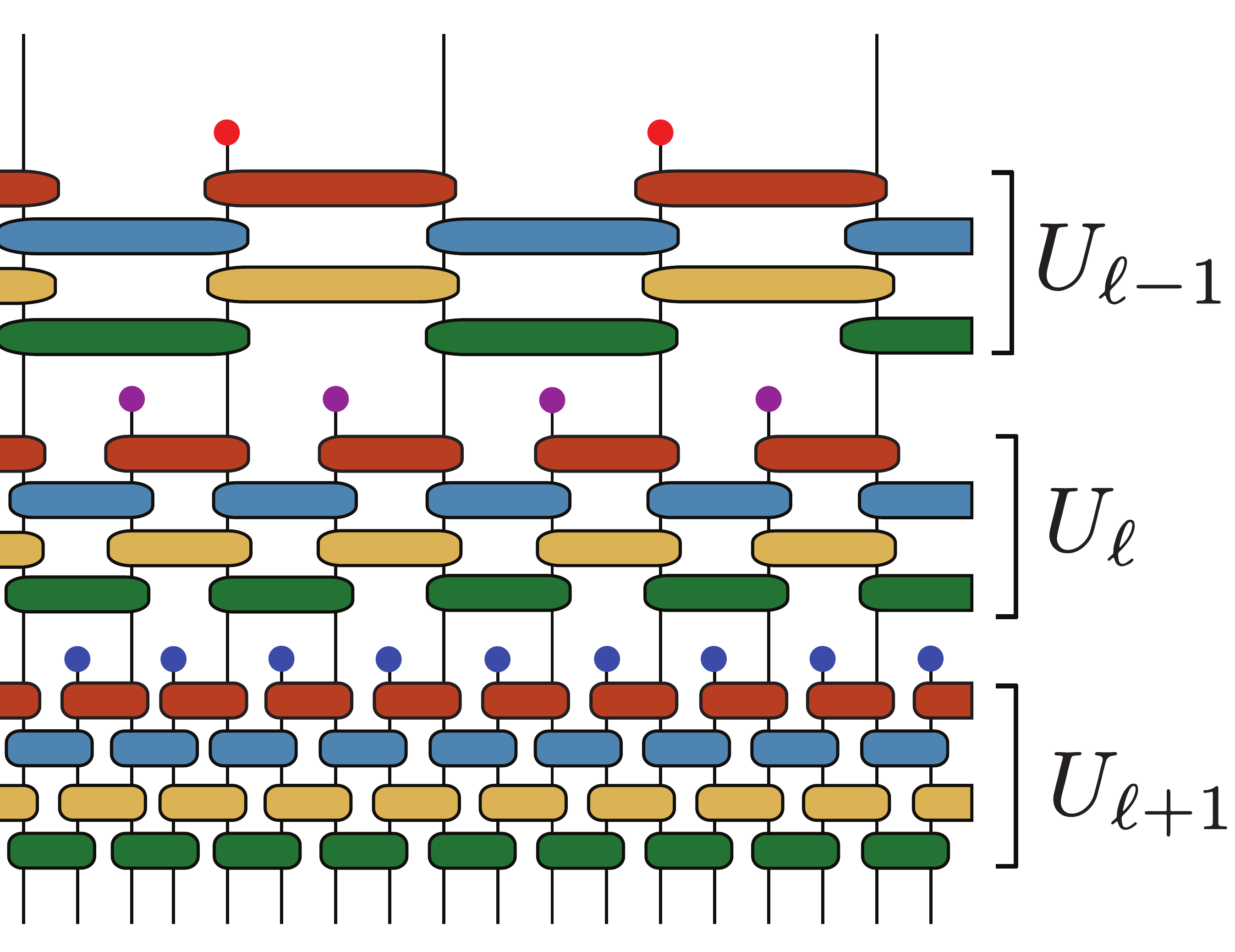}
   \caption{Partial circuit diagram for preparing  a TMERA state on a one dimensional lattice. Three scale transformations $\mathcal{S}_\ell$ implemented with local depth four brickwork circuits are shown, with time going from top to bottom in the direction of the UV scale. The gates are chosen to be identical in the different scale transformations to impose a discrete scale symmetry. Within each scale transformation the gates are also identical when translated by an even number of sites to impose a subgroup of translation symmetries. Mixed product states }
  \label{fig:TMERA}
 \end{figure}

A more expressive variant of the MERA can be constructed by
replacing the unitaries and isometries with low-depth quantum circuits.
This circuit ansatz is known as a \textit{deep MERA} (DMERA)\cite{DMERA}.
Like a MERA, a DMERA is built from successive scaling transformations.
But now the scaling transformation has two steps:
first, it adds $\ell$ new qubits interleaved between the $\ell$ qubits resulting from the previous layer
and initializes the new qubits in the $\ket 0$ state;
second, it applies a brickwork unitary circuit on the $2\ell$ qubits, entangling new with old.
Formally the scale transformation is 
\begin{equation}
\label{eq:dmera-scaling}
\ket{\psi_{\ell+1}} = U_\ell(\theta)\Big(\ket{ \psi_\ell} \otimes \ket 0^{\otimes 2^\ell}\Big)\;.
\end{equation}
Fig.~\ref{fig:TMERA} illustrates a closely related circuit (our TMERA, to be introduced below);
were that circuit a DMERA, all of the colored balls would represent the state $\ket 0$.

The DMERA is scale invariant because we require each of the scaling transformations to have identical parameters.
The total number of gates for preparing an $L$ qubit ansatz state is $O(D L )$,
$D$ the depth of the brickwork unitaries $U_{\ell}$,
so we might expect the number of variational parameters to likewise be $O(DL)$.
But translation symmetry reduces the count to $O(D \log(L))$ variational parameters,
proportional to overall circuit depth, 
and the scaling symmetry of critical systems further reduces the count to only $O(D)$ parameters%
---the depth of one scaling transformation.

Turn now to variational preparation of a thermal state. 
Our \textit{thermal multi-scale entanglement renormalization ansatz} (TMERA) 
prepares a mixed state by replacing the input states $\ket 0$ in the DMERA scaling transformation
with mixed product states $\sigma_{\beta,\ell}$;
the states $\sigma_{\beta,\ell}$ depending on system's inverse temperature $\beta$
and the scale $\ell$ at which the qubit is introduced.

To understand how this changes the state, imagine flipping a single one of the input states $\ket 0$ of Eq.~\ref{eq:dmera-scaling} at a single layer $\ell$ to $\ket 1$.
Once this change propagates to the physical qubits,
the result is an wavelet excitation at a length scale $L/2^{\ell}$,
where $L$ is the total length of the system.
The excitation introduced at layer $\ell$ therefore has an energy
\begin{equation}
    E_\ell \sim (2^{\ell}/ L)^z\;
\end{equation}
where $z$ is the model's dynamical critical exponent.

The DMERA circuit propagates excitations local on the input qubits to wavepacket excitations in the final state, which are semi-localized in both space and energy. The most energy-localized basis for a translation invariant system would be a plane-wave basis, i.e. the momentum modes of a free field theory,
but this transformation from a local to plane-wave basis is difficult to achieve using local quantum circuits. The DMERA circuit achieves a happy medium by being somewhat localized in energy, but still localized enough in space to be of short depth. 
The DMERA circuit can be viewed as implementing a kind of many-body quantum wavelet transform analogous to the Fourier transform between local and plane-wave bases. Wavelet transforms are used in classical signal processing for time-frequency analysis where localization in both time and frequency are required.

The energy localization of the wavelet basis allows us to use the local input qubits as a good basis to use for approximately filling out the distribution of our target Gibbs state and motivates the use of a product spectrum ansatz.
Since initial scales are associated with low energy excitations, they have the largest probability of being excited in a thermal state, with successive layers having smaller probabilities dictated by the energy localization of that scale. All qubits introduced at a given scale will have the same excitation probability because they are all related by translations which leave the circuit invariant below that scale.

\begin{figure}
    \centering
   \begin{minipage}{\linewidth}
  \includegraphics[width=\linewidth]{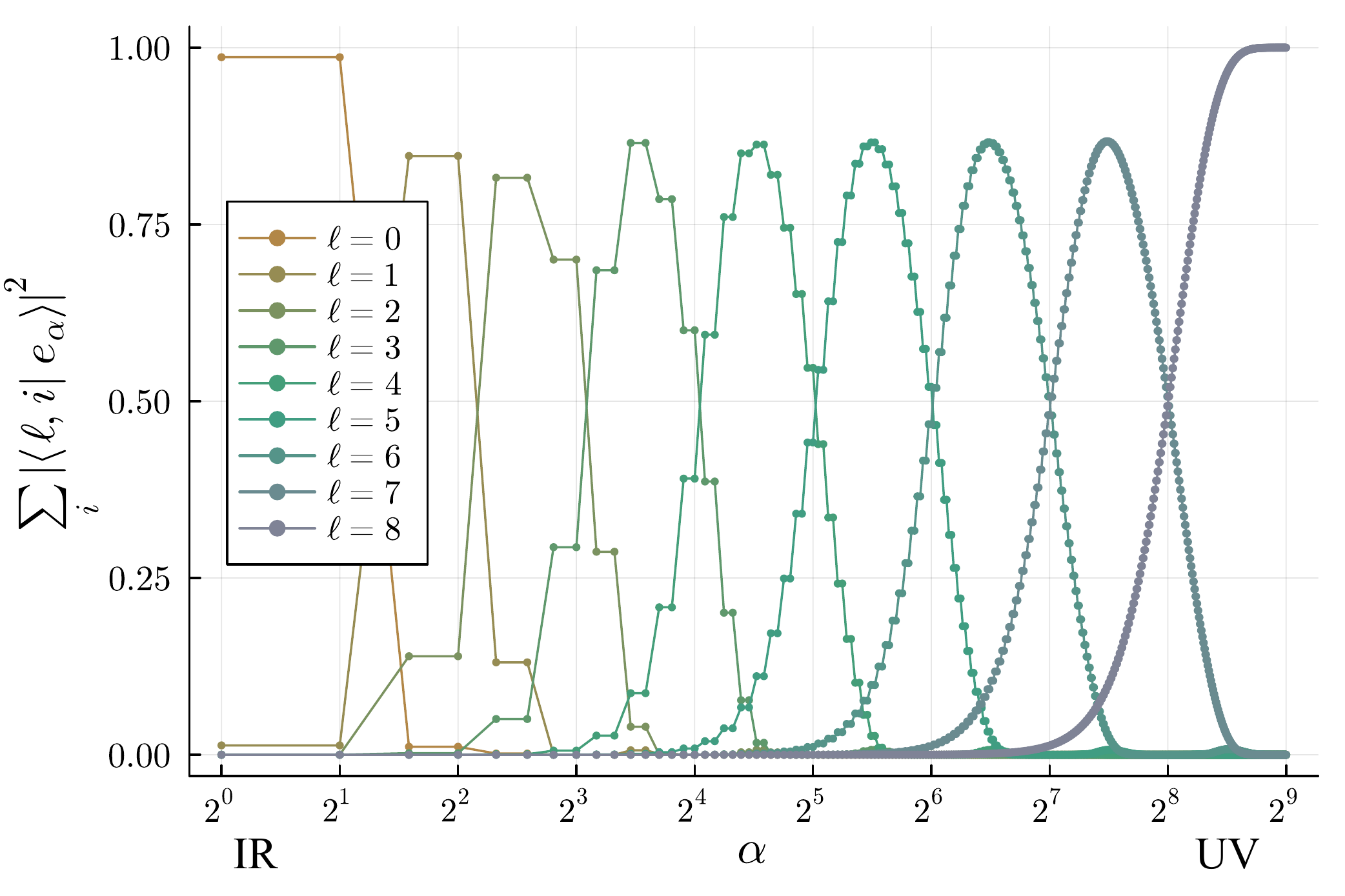}
 \end{minipage}
   \caption{The support of the first-quantized wavepacket modes $| \ell, i \rangle$ introduced at layer $\ell$ with the energy eigenmodes $\ket{e_\alpha}$. The inner product is summed over all $2^\ell$ from each layer, showing the support for each scale is localized in energy. }
  \label{fig:waveSupp}
 \end{figure}

In Fig. \ref{fig:waveSupp} we see that the localized wave packet modes introduced in layer $\ell$ of our circuit are supported mostly by energy eigenmodes $e_\alpha$ within a narrow energy window.

Varying the energies $E_\ell$, and
hence the mixed states $\sigma_\ell(\beta)$,
breaks the scale invariance of the circuit ansatz. 
This reflects the physics of the Gibbs state:
the temperature introduces a characteristic scale.

\subsection{Circuit ansatz}
Precisely, then, the thermal multi-scale entanglement renormalization ansatz (TMERA) prepares an approximate thermal state $\rho_L(\beta)$ on $2^L$ qubits at inverse temperature $\beta$
in a scale-by-scale manner
by preparing a state $\rho_\ell(\beta)$ on $2^\ell$ qubits for each scale $\ell$.
Each scale transformation $\mathcal{S}_\ell$ prepares the state $\rho_\ell(\beta)$ from $\rho_{\ell-1}(\beta)$
by applying a unitary circuit $U_\ell(\theta)$
which entangles the prior state $\rho_{\ell-1}(\beta)$
with $2^{\ell-1}$ new qubits,
each introduced in the single qubit mixed state $\sigma_{\beta,\ell}$.
The scale transformation is therefore
\begin{align}
  \label{eq:tmera-scaling}
  \begin{split}
   \rho_{\beta,\ell} &\equiv \mathcal{S}_{\beta,\ell} [\rho_{\beta,\ell-1}]\\
   &= U_\ell (\theta) ( \rho_{\beta, \ell -1} \otimes \sigma_{\beta, \ell-1}^{\otimes n} ) U^\dagger_\ell(\theta)\;.
  \end{split}
\end{align}
We illustrate three applications of the scale transformation in Fig.~\ref{fig:TMERA}.
The balls represent $\sigma_{\beta,\ell}$, and the rounded rectangles the gates of the brickwork circuit $U_\ell$.

The mixed state $\sigma_{\ell}(\beta)$ is a Gibbs state of of the Pauli matrix $Z$
\begin{align}
  \label{eq:single-gibbs}
  \begin{split}
   \sigma_{\beta,\ell} = \mathrm{sech}(\beta E_\ell) e^{ \beta E_\ell Z} \;.
  \end{split}
\end{align}
We can choose $\sigma_{\beta,\ell}$ in the form \eqref{eq:single-gibbs} without loss of generality,
because unitaries on the qubit can be absorbed into the brickwork circuit $U_\ell$.
These density matrices are easily prepared either by taking an ensemble over classical randomness, or by taking one of two entangled qubits.

We imagine two families of variational circuit,
depending on what is known in advance about the model.
If  the dynamical critical exponent $z$ is known,
    then the scale transformations $U_\ell$ are those of the DMERA,
    and the energies are
    \begin{align}
    \label{eq:energy-scaling}
    E_\ell = E_{UV} (2^{\ell}/L)^z
    \end{align}
    ---with the single variational parameter $E_\UV$.
If the dynamical critical exponent is \textit{not} known, 
    the energies $E_\ell$ are independent variational parameters from which it can be extracted.
(If the ground state DMERA is not known, it can be found by a separate pure state variational optimization.)
We focus on the case in which the scaling exponent is known,
but we verify that it can be extracted from independent optimization of the $E_\ell$.

Our ansatz also naturally gives an algorithm for preparing approximations to the purification of the thermal state known as the thermofield double (TFD) state.
The TFD state is a pure state on two copies of the system, where the reduced density matrix on either copy looks like a Gibbs state at a given temperature. 
In our case, the qubits introduce in mixed product states for one copy would instead become pairs of entangled qubits for preparing an approximate TFD state. Identical unitary circuits would then be applied to each collection of entangled qubits so that two copies of the approximate Gibbs state are prepared which purify each other. The circuit diagram at the top of Fig. \ref{fig:highlights} can be doubly interpreted as applying a unitary circuit to a product state density matrix as is the case for Gibbs state preparation, or applying two unitary circuits to each half of a product of entangled pairs as in the case for TFD state preparation. 

\section{Model}\label{s:model}

We benchmark the TMERA on the one dimensional critical transverse-field Ising model
\begin{align}
  \label{eq:spin-ham}
  \begin{split}
    H_I &=  - \sum_j \sigma^x_j \sigma^x_{j+1} + \sigma^z_j\;.
  \end{split}
\end{align}
At low energy this spin chain is known to be well described by a conformal field theory with $c=1/2$; \cite{DiFrancesco1997}
its ground state is well approximated by a DMERA circuit.\cite{ERwav, PhysRevX.8.011003}

The transverse field Ising model is free fermion integrable.
The Jordan-Wigner transformation maps the Hamiltonian \eqref{eq:spin-ham} to the quadratic Majorana fermion Hamiltonian
\begin{align}
  \label{eq:f-ham}
  \begin{split}
    H_f &=   i \sum_{j=1}^{2L} \gamma_j \gamma_{j+1}
  \end{split}
\end{align}
Although spin chain models like \eqref{eq:spin-ham} are natural for digital quantum computers,
free fermion models like \eqref{eq:f-ham} are more natural for classical computers.
Many states of interest%
---including the Gibbs states with which we are concerned---%
are Gaussian fermion states.
They are therefore (at least in principle) preparable by matchgate circuits,
which are equivalent to unitary Gaussian operations.

Covariance matrix techniques can simulate large matchgate circuits efficiently.
For models that are not free fermion,
the target Gibbs state will no longer be Gaussian,
and the ansatz circuit must include gates outside the set of matchgates.
This is not an obstacle for a quantum computer,
but it is an obstacle for a classical computer%
---which is why we restrict ourselves to the free-fermion Hamiltonian \eqref{eq:f-ham}.

Matchgates map single-fermion states to single-fermion states.
Since the DMERA (and TMERA) for a free fermion model like \eqref{eq:f-ham} consists entirely of matchgates,
we can rephrase the wavelet excitation picture of Sec.~\ref{s:ansatz-intuition} even more explicitly.
The matchgate DMERA circuits transform local fermion modes introduced at scale $\ell$ into wavepacket modes with support over $2^\ell$ lattice sites in the final state.
This linear transformation on the modes is equivalent to a discrete wavelet transform.
A Gaussian fermion state may be specified by a set of fermion operators which annihilate it, and any linear combination of these operators also annihilate the ground state. A Gaussian DMERA circuit ansatz may be understood as preparing the state annihilated by these wave packet modes introduced at each scale, which span approximately the same subspace of operators as the actual ground state annihilation operators.
Analogously to how periodic functions may have a simple representation in a Fourier basis, the success of a DMERA circuit ansatz lies in the ability to find a good wavelet basis that admits a simple approximation of the target ground state with acceptable fidelity. 

While this clear picture in terms of free fermion modes breaks down for non-integrable models, we may still think of the DMERA circuit ansatz with a universal gate set as a more general kind of many-body wavelet transformation. This ansatz will still have a product spectrum, but may still be successful as long as the target state has an entanglement structure that admits this scale-by-scale coupling of degrees of freedom. 

The underlying DMERA circuits for ground state preparation that we use are from \cite{gsvar_dmera}, where circuit parameters were optimized using the expectation value of the energy density. The energy density was optimized using the fixed-point density matrix of the local quantum channel that implements a DMERA scaling circuit on a constant number of qubits, effectively optimizing the average energy density of the infinite volume anstz state. These parameters are then used to prepare 

This parameterized family of circuits is a modest generalization of the circuits proposed in \cite{ERwav} to prepare the Ising model ground state, with each local two-qubit gate being the real-valued matchgate $u(x,y)$ that implements independent rotations of the even- and odd-parity subspaces
\begin{equation}
\label{eq:local-gate}
u(x,y) = \begin{bmatrix}
\cos(x) & 0 & 0 & \sin(x) \\
 0& \cos(y) & \sin(y) & 0 \\
 0& -\sin(y) & \cos(y) & 0 \\
 -\sin(x)& 0 & 0 & \cos(x) \\
\end{bmatrix}\;.
\end{equation}

So, the DMERA circuits have $2D$ parameters in total after imposing constraints for approximate scale and translation invariance expected in a critical ground state.

\section{Free energy minimization and properties of the TMERA}\label{s:properties}


Variational optimization of the Gibbs state at temperature $T$ requires minimizing the free energy
\begin{equation}
\label{eq:free-energy}
 F = E - TS
\end{equation}
over our variational parameters.
This in turn requires measuring energy and entropy each Gibbs state.
While the energy of the ansatz state must be measured from the expectation values of local terms of the Hamiltonian,
the entropy can be inferred directly from the excitation probabilities of wavelet modes. 

\subsection{Entropy}\label{ss:entropy}

A TMERA prepares a mixed state by repeated application of the scaling transformation
\begin{align}
  \begin{split}
  \mathcal{S}_{\beta,\ell} [\rho_{\beta,\ell-1}]
  &= U_\ell (\theta) ( \rho_{\beta, \ell -1} \otimes \sigma_{\beta, \ell-1}^{\otimes n} ) U^\dagger_\ell(\theta)\;.
  \end{split}
\end{align}
Consequently the final mixed state is---up to a unitary---a tensor product of the mode mixed states $\sigma_{\beta, \ell - 1}$,
and the entropy is the sum of the mode entropies.
Because each scale has $2^{\ell}$ modes in total, the total entropy is
\begin{align}
  \label{eq:ansatzEnt}
  \begin{split}
  S = \sum_{\ell=1}^n 2^{\ell} S_\ell .
    \end{split}
\end{align}
where
\begin{align}
  \label{eq:modeEnt}
  \begin{split}
  S_\ell &=  -p_\ell \log ( p_\ell) -(1 - p_\ell) \log (1 -p_\ell) \\
  &= \log( \cosh (\beta E_\ell) ) - \beta E_\ell \tanh( \beta E_\ell ) + \log 2. \\
    \end{split}
\end{align}
is the entropy of a single qubit in the state $\rho = Z^{-1} e^{-\beta E_\ell}$\;.

\begin{figure}[t]
  \centering
    \includegraphics[width=0.45\textwidth]{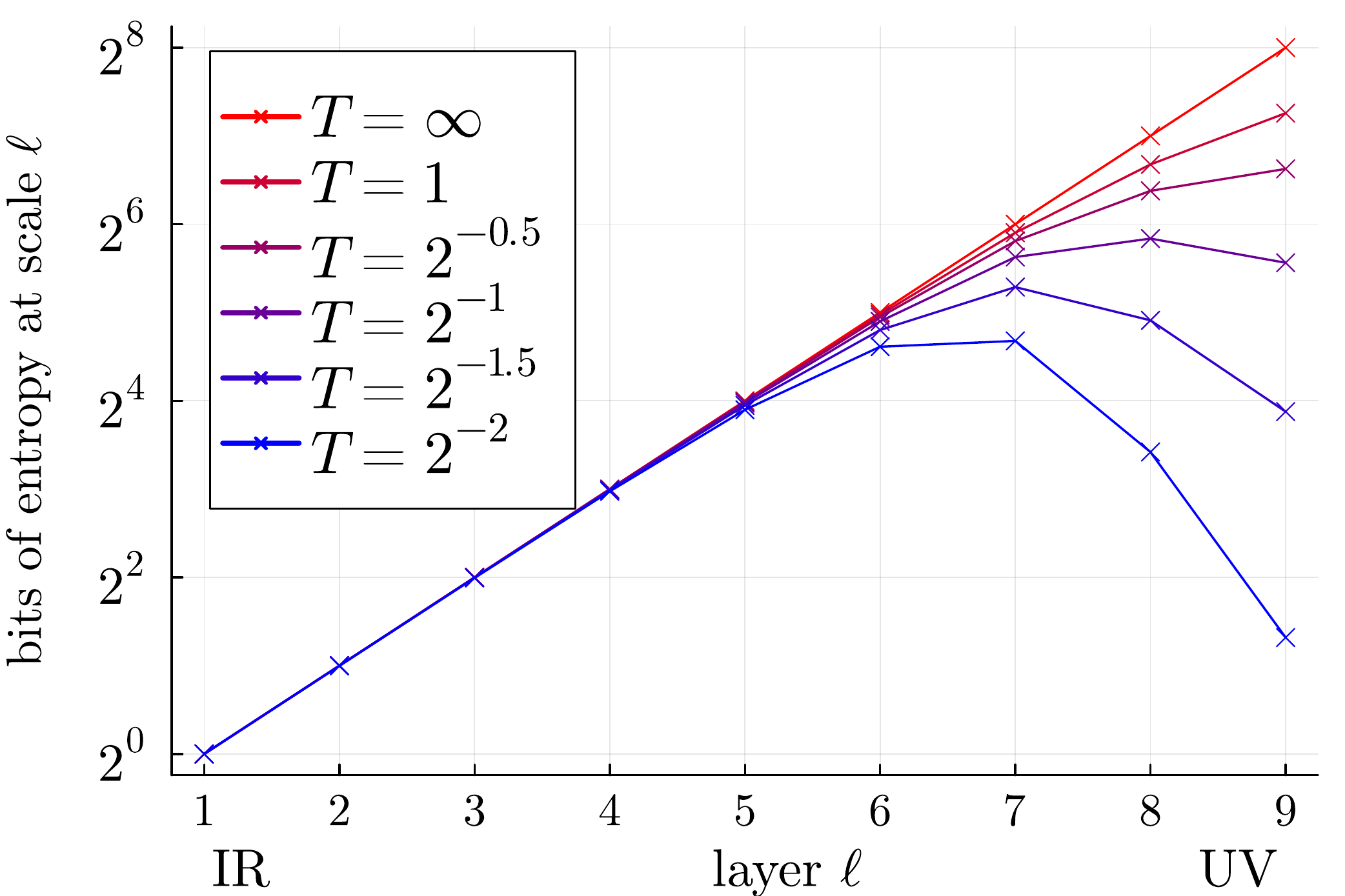}
  \caption{Entropy introduced at each scale. At high temperature most of the entropy comes from the many UV modes near infinite temperature. At low temperature a non-monotonicity results from competition between the number modes at each scale and the entropy per mode . }
 \label{fig:scaleEnts}
\end{figure}

In variationally optimizing the free energy \eqref{eq:free-energy},
one would explicitly evaluate the full expressions \eqref{eq:ansatzEnt} and \eqref{eq:modeEnt}.
But we can get physical insight into the origin of the state's entropy
by considering the UV and IR limits of \eqref{eq:ansatzEnt} and \eqref{eq:modeEnt}.
In those limits, the entropy of each mode is
\begin{subequations}\label{eq:Sell-limits}
\begin{alignat}{2}
        &S_\ell \approx \log(2) - \frac{1}{2}(\beta E_\ell)^2&&\qquad \beta E_\ell \ll 1 \text{ (IR)}\\
        &S_\ell \approx 2\beta E_\ell e^{-2 \beta E_\ell} &&\qquad \beta E_\ell \gg 1 \text{ (UV)}\;;
\end{alignat}
\end{subequations}
an the energy is $E_\ell \propto 2^{\ell z}$. 
Comparing these entropies with the total entropy of Eq.~\eqref{eq:ansatzEnt},
we see a competition between the number of modes and the entropy per mode.
In the UV limit the entropy per mode may be exponentially small, but there are exponentially many modes.
In the IR limit, by contrast the individual mode entropies approach their maximal value of $\log(2)$,
but there are $O(1)$ modes.
In Fig. \ref{fig:scaleEnts} we show the total entropy contribution from each scale at a variety of temperatures. 
For high temperatures all modes are close to maximally mixed, so the entropy contribution follows the exponential decay of the number of modes at each scale. 
For low temperatures, however, the UV modes are relatively unexcited and do not contribute much entropy despite the large number of them, and the bulk of the total entropy comes from scales with energy low enough to to have non-negligible excitation probability.
A given scale $\ell$ will contribute more entropy than the shorter length scale $\ell+1$ when $S_\ell /2 > S_{\ell+1}$, which occurs when $\beta E_\ell \approx 1.26$.

\subsection{Energy}

Since the system is translation-invariant,
the energy can be computed by measuring the energy density on a single bond.
Such a measurement does not require the full DMERA:
rather, it requires the past causal cone of the single bond.
The past causal cone of a subsystem converges to a width independent of subsystem size or total system size, determined only by the scaling transformation depth $D$. With the ability to reset and reuse qubits, at least $4D -2$ qubits are needed to implement the this local depth $D$ channel that prepares a subregion of the state. 

\begin{figure}[t]
  \centering
    \includegraphics[width=0.45\textwidth]{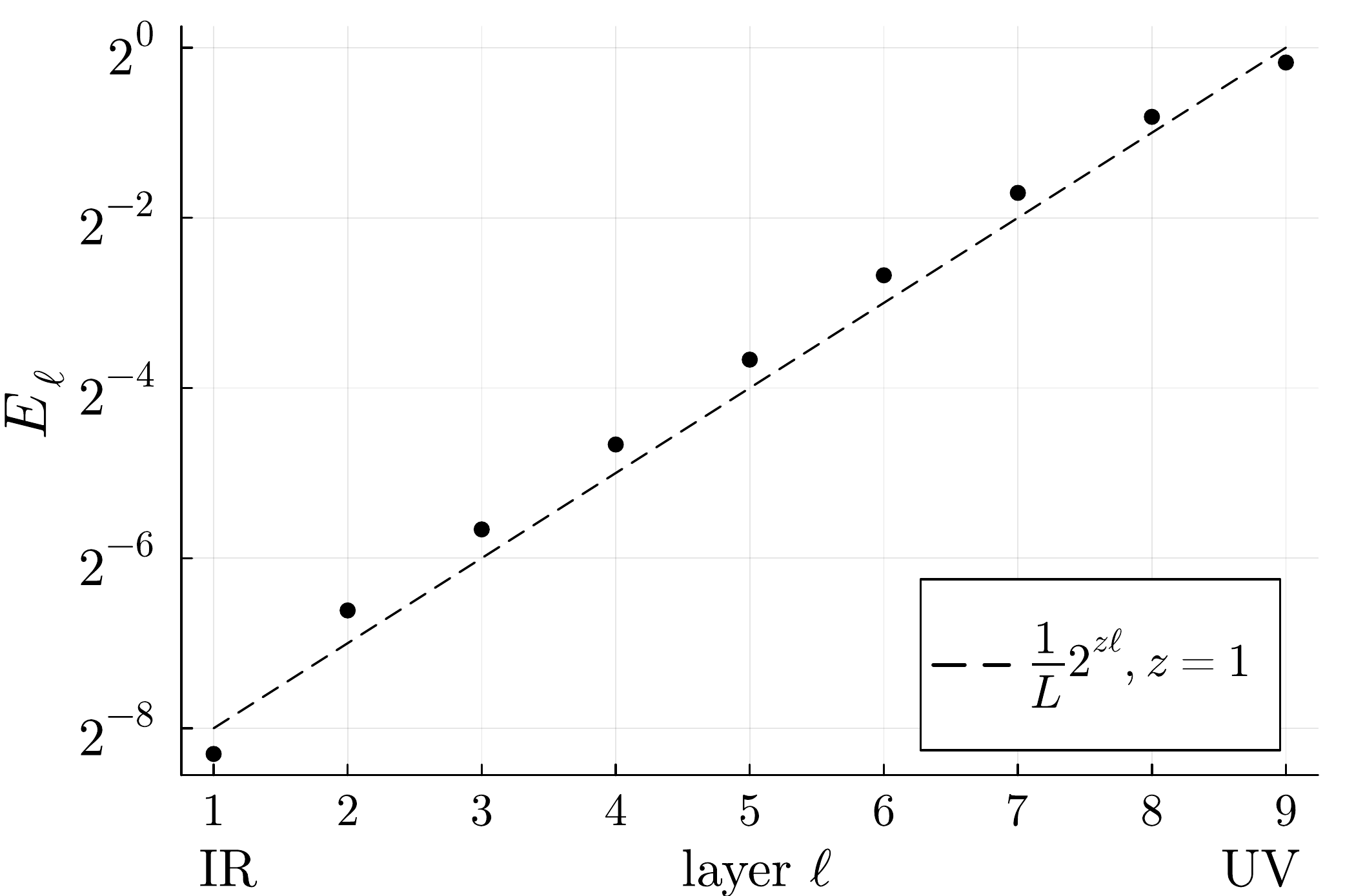}
    \caption{Average energy $E_\ell$ of modes excited at the $\ell^\mathrm{th}$ scale, with dashed line showing an exponential scaling proportional to $2^{z\ell}$ with $z=1$ for reference, system size $L=512$.
}
\label{fig:Escale}
\end{figure}

We are benchmarking on a free-fermion integrable model with a matchgate DMERA,
so the entropy-maximizing energy is just the mode expectation value of the physical Hamiltonian
(cf. App.~\ref{app:psa-mean-field}).
In Fig.~\ref{fig:Escale} we plot those mode energies,
and we see that the mode energies broadly follow the expected scaling $E_\ell \propto (2^{\ell})^z$.
They depart from that scaling at the lowest $\ell$ (IR-most modes),
where finite size effects become important,
and at the highest $\ell$ (UV-most modes), where lattice effects become important.

\section{Benchmarking}\label{s:benchmarking}


In Sec.~\ref{s:properties} we described how to variationally optimize a TMERA.
In this section we measure how well the resulting state approximates the Gibbs state.

\subsection{Energy and Entropy}

One measure of accuracy of our TMERA states is observing the relationships between energy, entropy, and temperature. One way of defining a state in thermal equilibrium is that it is the density matrix of highest entropy with a given expectation value of the energy, or equivalently the lowest energy state with a given entropy. The solid black line in Fig. \ref{fig:highlights} traces this extremal relation between energy and entropy which is occupied by positive temperature Gibbs states. All other quantum states must lie in the region below this extremal line, so the degree to which our thermal ansatz states can hug this line is a measure of their success. 

For a state with $L=512$, the largest difference in entropy for the TMERA state with $D=6$ is around 0.014 bits per qubit and occurs near temperature $T=0.28$, while the $D=1$ state has the largest entropy difference of around 0.06 bits per qubit near $T=0.19$. The largest energy differences are 0.011 per qubit for $D=6$ near $T=1.3$ and 0.096 per qubit for $D=1$ near temperature $T=0.87$.

\begin{figure}[t]
   \begin{minipage}{0.75\linewidth}
  \includegraphics[width=\linewidth]{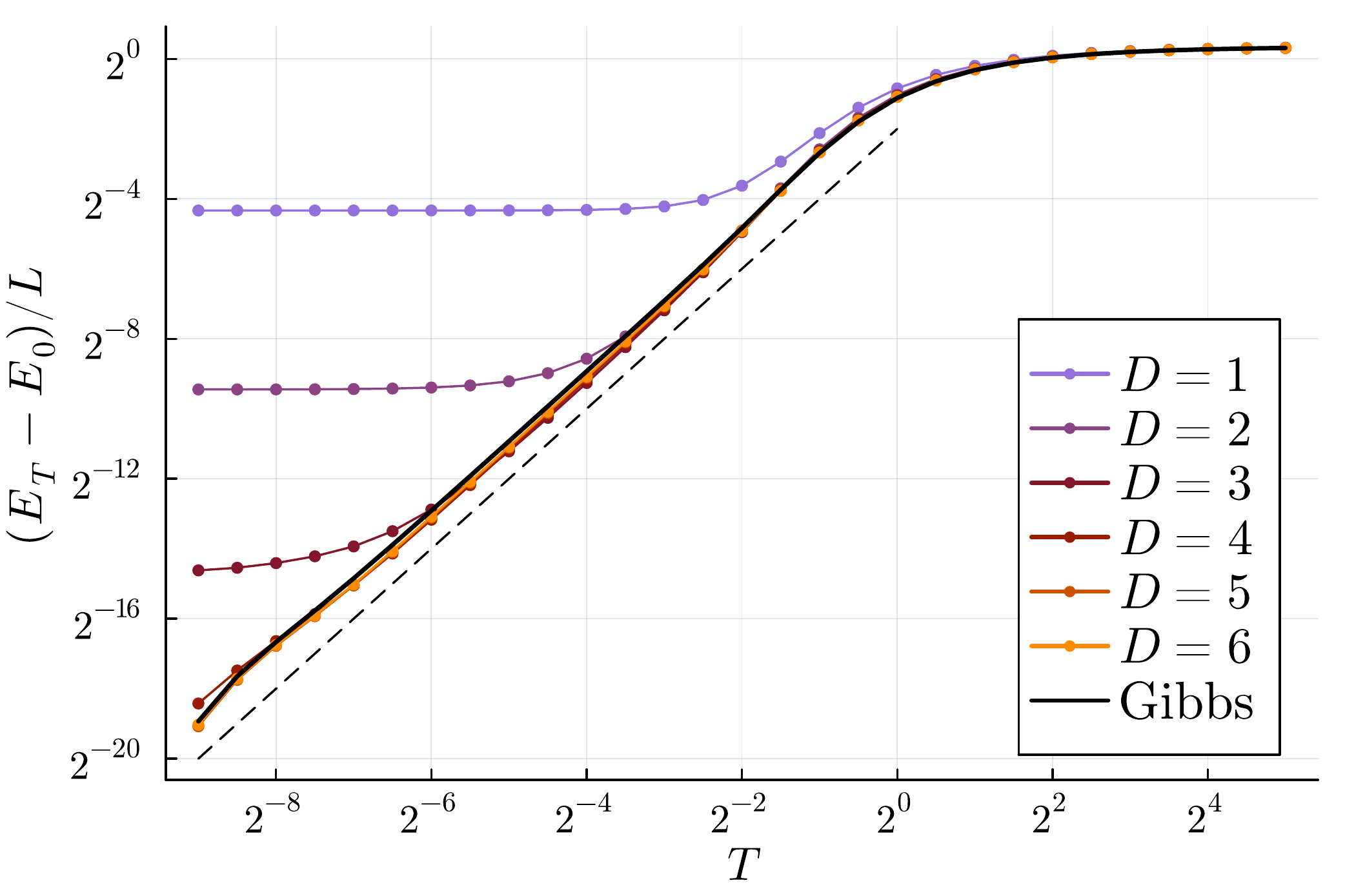}
 \end{minipage}\hfill
   \begin{minipage}{0.75\linewidth}
  \includegraphics[width=\linewidth]{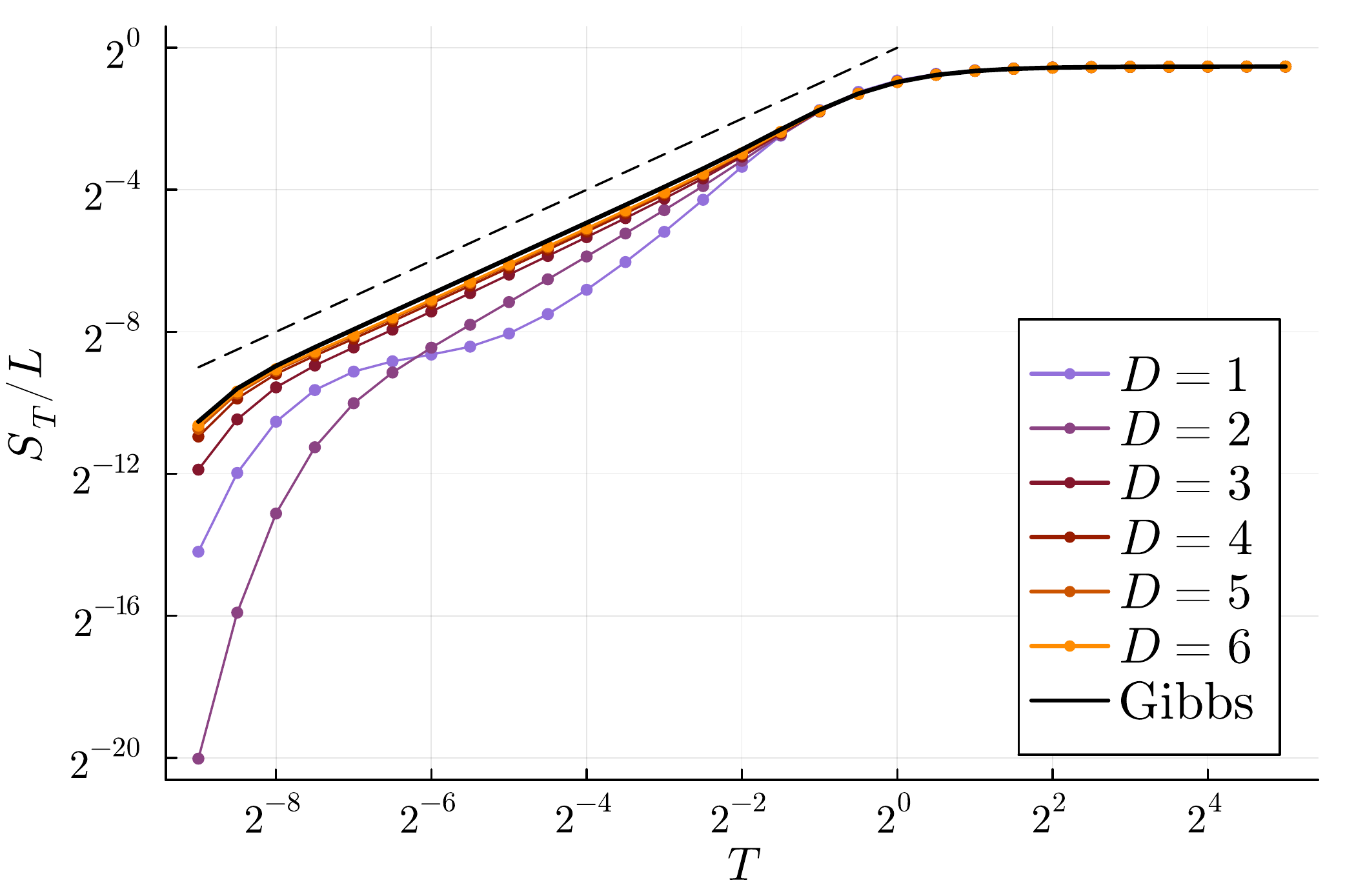}
 \end{minipage}
  \caption{\textbf{Top:} Log-log plot of the energy above the ground state energy per qubit vs temperature, with dashed line showing a quadratic scaling for reference.
  \textbf{Bottom:} Log-log plot of the entropy per qubit vs temperature, with dashed line showing a linear scaling for reference.
  }
  \label{fig:S-E-scaling}
 \end{figure}

For low temperatures we observe polynomial scaling of entropy and energy with temperature, as seen in Fig. \ref{fig:S-E-scaling}. The log-log plots seem consistent with linear and quadratic scaling for the entropy and energy respectively. This is exactly what we expect from the long-wavelength description in terms of a conformal field theory. In any such theory, the scaling symmetry implies that the temperature sets a thermal length scale $\xi(T)\sim 1/T$ (with $k_B$, $\hbar$, and the ``speed of light'' set to unity). As the only dimensionful quantity in the problem, this length scale determines the temperature dependence of the entropy density, $s \sim \xi^{-d}$, and the energy density, $\epsilon \sim \xi^{-(d+1)}\sim T \xi^{-d}$, where $d$ is the dimension of space. Setting $d=1$ gives $s \sim T$ and $\epsilon \sim T^2$, consistent with the data. Note that the shorter depth ansatz states deviate at low temperature when they cannot lower the energy beyond the ground state ansatz for depth $D$.

\subsection{State Fidelity}

\begin{figure}
  \centering
  \begin{minipage}{0.45\textwidth}
  \includegraphics[width=\linewidth]{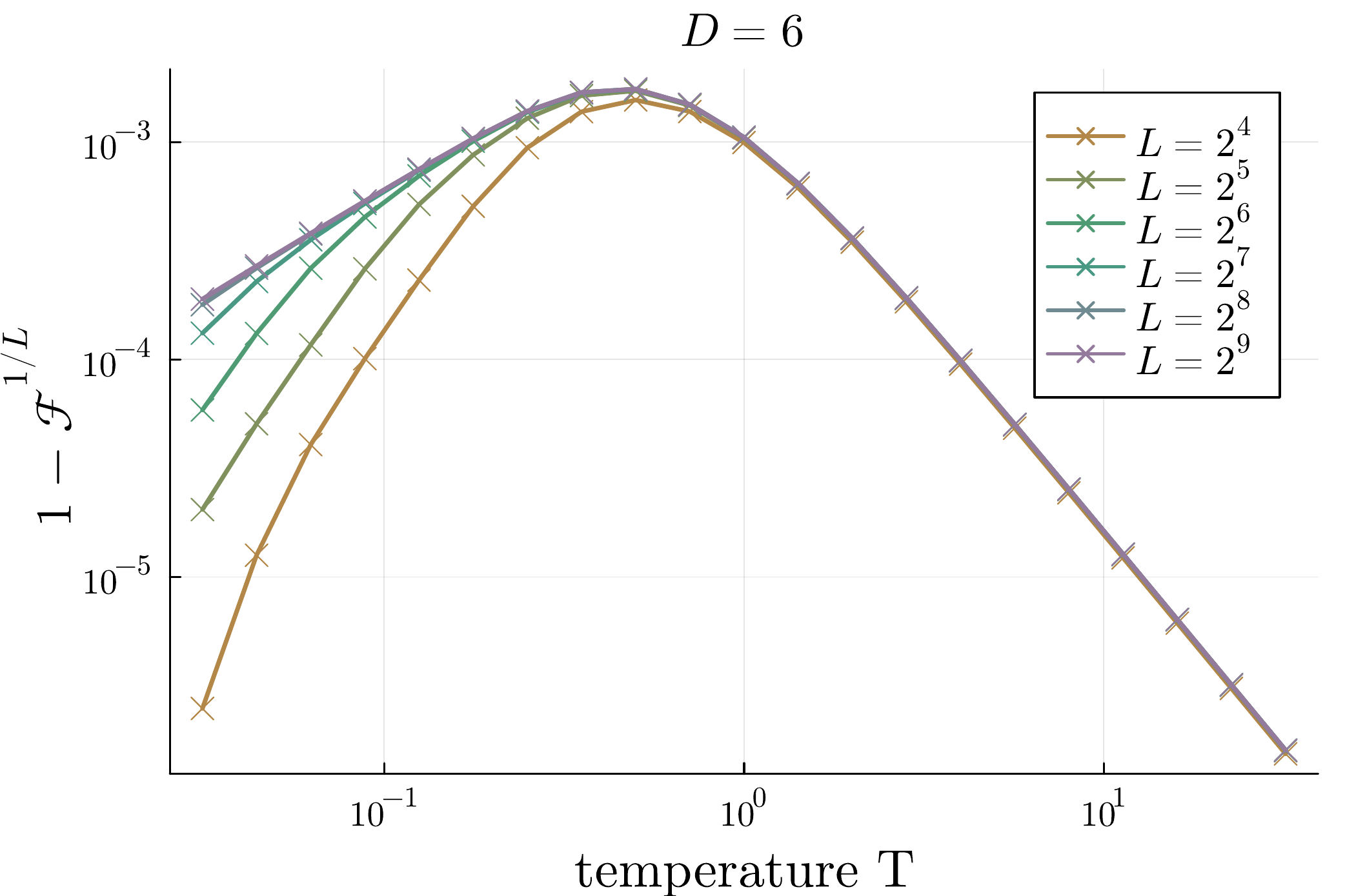}
  \end{minipage}
     \caption{
     Scaled fidelity for $D=6$ TMERA states for various temperatures $T$ and system sizes $L$. 
  }
 \label{fig:ThermFidBT}
\end{figure}

Despite the simplicity of the TMERA ansatz, our approximate state still has high fidelity for $L=512$: the fidelity is $\mathcal{F} > 0.4$ for all temperatures for the $D=6$ states \ref{fig:highlights}. Even the $D=3$ ansatz states have fidelity higher than 0.2 for all temperatures. 
The fidelity scales as
\begin{align}
    1 - \mathcal{F}^{1/L} = s(T),\qquad s(T) < 10^{-2}\;.
\end{align} 
(Fig.~\ref{fig:ThermFidBT})
For high temperatures the fidelity displays this scaling even for the smallest system sizes we treat ($L = 16$);
for low temperatures the fidelity approaches this scaling for $L \gtrsim 128$.

While ground state fidelity increases exponentially with depth, thermal state infidelity per site does not seem to decrease much further than $2^{-10}$ for most finite temperature states. This could be due to some limitations of the product spectrum of our approximate density matrix which is one inherent limitation to the expressiveness of our ansatz as is.

\subsection{Correlation Function and effective Hamiltonian}

The ground state of a gapless system in one dimension has a quadratic correlation function $\langle \gamma_i \gamma_j \rangle$ which decays polynomially with distance as $\sim |i-j|^{-1}$. The thermal energy scale breaks scale symmetry at finite temperature and and introduces a correlation length, leading to exponential decay of this correlation function, with $\log ( \langle \gamma_i \gamma_j \rangle_\beta ) \sim -\beta |i-j|$. 

The TMERA states, however, maintain polynomially decaying quadratic fermion operators, albeit with$\langle \gamma_i \gamma_j \rangle \sim |i-j|^{-2}$. Despite this very different behavior of the correlation function over long distances, our approximate Gibbs states still have relatively good fidelity. These values of these long distance observables are exponentially small, so despite greatly overestimating them proportionally by using a polynomially decaying function, the error is still within a small additive precision and does not impact the overall fidelity too greatly.

\section{Discussion}

We propose a variational quantum circuit ansatz for preparing approximate Gibbs states called a \textit{thermal multiscale renormalization ansatz}, or TMERA.
Our ansatz targets states in which scale invariance is broken only by temperature, as in a quantum critical fan.
The TMERA builds on a deep multiscale renormalization ansatz (DMERA) for the ground state, which we presume is already known
(perhaps via an independent variational optimization).
For a 1d system of length $L$,
the TMERA has $O(\log L)$ parameters if the system's dynamical critical exponent is not known;
if that exponent is known, it has a single parameter.
The TMERA requires $O(DL)$ gates, $D \sim 1-10$ the depth of a brickwork unitary circuit used in scaling.

In a TMERA both measurements required for optimizing free-energy---energy and entropy---are natural.
The entropy follows directly from the structure of the TMERA and the variational parameters.
The energy can be measured using only the past causal cone of a two-site region;
this past causal cone is a subset of the full TMERA involving only $O(D\log L)$ gates.

We benchmarked our TMERA on the free-fermion integrable transverse-field Ising model at its critical point.
We found that with a $D=6$ circuit the ansatz produced fidelity $\mathcal F \gtrsim 0.4$ on a 512-site system,
per-site infidelity $1 - \mathcal F^{1/L} < 10^{-2}$ for all system sizes and temperatures,
and nearly maximal entropy at fixed energy.

Neither the basic inputs nor the physical intuition behind the TMERA are specific to the transverse field Ising model we used for bechmarking.
The inputs are a scale-invariant DMERA network describing the ground state and a scaling ansatz for the characteristic energy as a function of scale;
these inputs reify the intuition that ground states of models at critical points are scale-invariant,
and thermal states are governed by the model's dynamical scaling.
A critical point is described at long wavelengths by a conformal field theory
(or a more general non-relativistic analog)
and the scaling symmetry of the long-wavelength theory is expected to imply the existence of analogous structures independent of the details of the theory.
In other words, our construction can plausibly apply to a wide variety of quantum critical points in diverse dimensions. It may of course be challenging to classically simulate the resulting tensor networks, especially in higher dimensions, but our architecture would remain a potentially powerful quantum variational ansatz that leverages our significant a priori expectations for the physics.

\begin{figure}
  \centering
    \includegraphics[width=0.45\textwidth]{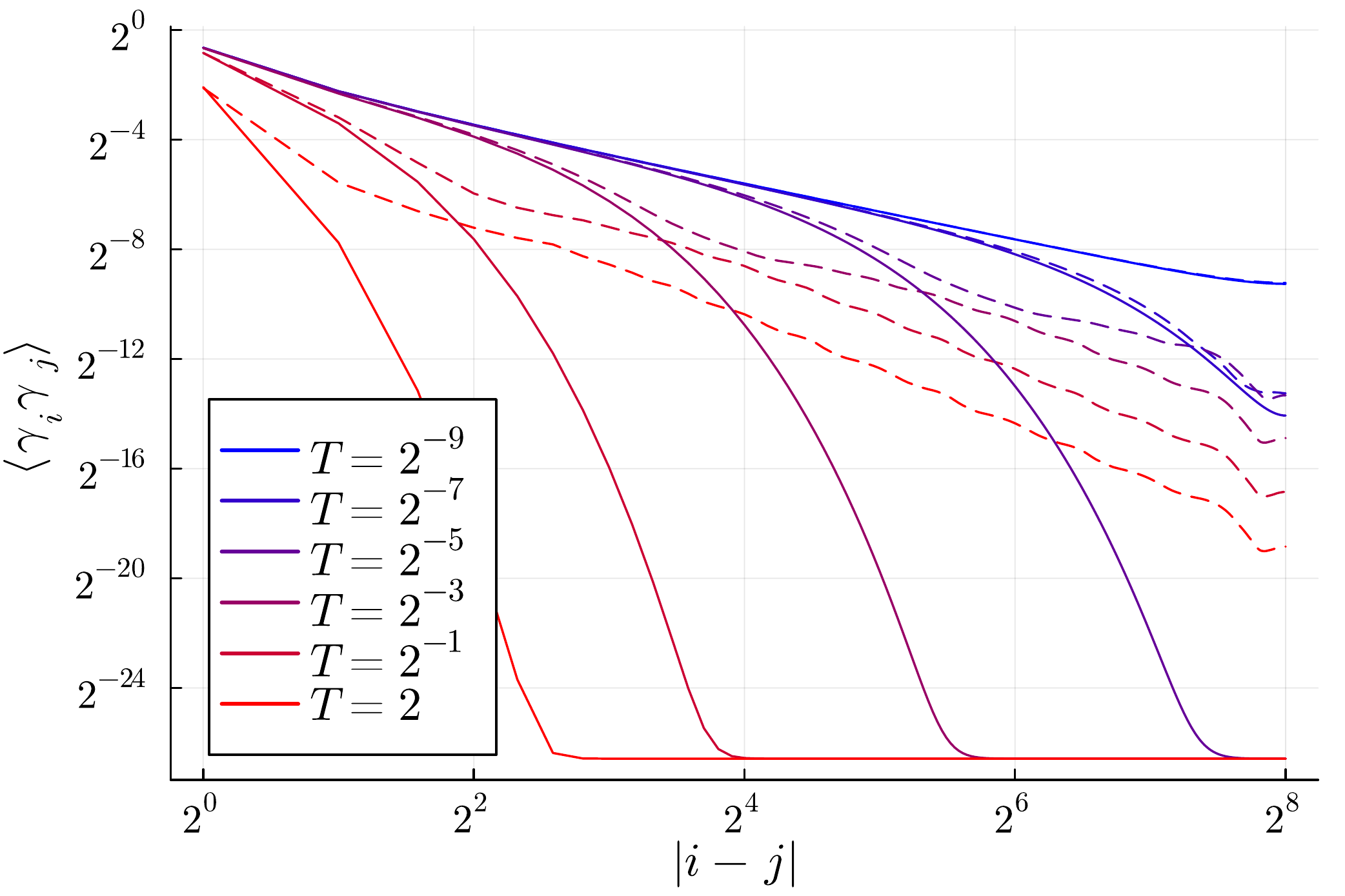}
  \caption{Two-fermion correlation functions, featuring exponential decay for finite temperature Gibbs states and polynomial decay for TMERA states, with $D=6$ and $L=512$. The lowest temperature of $T=1/L$ converges to the highly accurate ground state correlation function due to finite size, while higher temperature correlation functions only match well for larger expectation values, i.e. those within a correlation length set by the temperature.}
 \label{fig:2ptf}
\end{figure}

\subsection{Interacting systems}

In App.~\ref{app:psa-mean-field} we show that any product spectrum ansatz prepares the Gibbs state of a mean-field approximation to a classical stat mech model on the modes.
For free-fermion models,
like the transverse-field Ising model \eqref{eq:f-ham} we have used for benchmarking,
the mean-field approximation becomes exact.
For interacting models the mean-field approximation is no longer exact%
---but it is still justified.
Purely on structural grounds,
one expects the DMERA to transform local interactions between physical degrees of freedom
to non-local interactions between modes.
Fig.~9 confirms this expectation:
since the modes have power law correlations,
local operators will have  matrix elements between widely separated modes.
Provided the power is not too high, these interactions can be accurately treated in the mean-field interaction.
Indeed \cite{Martyn_2019} found that a product spectrum ansatz
(albeit with a different unitary structure)
can accurately describe the transverse field Ising model with integrability-breaking interactions,
and an interacting SYK model.

We can systematically improve the TMERA architecture by going beyond this mean-field approximation.
Although the TMERA is a product spectrum ansatz, in which the thermal entropy is injected via product states, the key feature we need for a thermal variational ansatz is the ability to estimate the free energy. This property holds for a much wider class of architectures, including those that allow local entanglement via multi-site thermal states and Markov states. This additional freedom, combined with the ability to vary the depth of the layers, gives considerable power to manipulate both the spectrum and the diagonal basis of the resulting density matrix. Given how well the product spectrum ansatz already performs, these extensions should be investigated as potential routes to an architecture that is systematically improvable.

\subsection{Extension to systems away from criticality}

For a critical ground state, the underlying ground state circuit is scale-invariant. The temperature imposes an energy scale $E_{UV}$ that breaks scale symmetry in the TMERA via the different mode energies at each scale $E_\ell \approx E_{UV} 2^{-\ell z}/L$.
If we tune away from the critical point, perhaps by changing a parameter $h$ to $h \ne h_c$,
we introduce another length scale $\xi \sim (h - h_c)^{-\nu}$ and energy scale $\xi^{-z}$.
The ground state is no longer scale invariant,
so one might worry that the TMERA is useless, or at least no improvement on a brickwork circuit.
But the system is locally scale invariant for lengths $l \lesssim (h - h_c)^{-\nu}$.
So the brickwork circuit will be redundant, perhaps highly so.

We believe a TMERA modified to take into account the scale $\xi \sim (h - h_c)^{-\nu}$ will eliminate the redundancy.
The correct TMERA will be cut off at 
\begin{align}
    n(h - h_c) \sim \log \xi \sim -\nu \log(h - h_c)
\end{align}
layers. 
This number of layers is not a priori known%
---but learning it is a matter for ground-state DMERA optimization,
not TMERA optimization.
Given the ground state DMERA, 
the TMERA construction and optimization will proceed as we have described here;
the resulting TMERA will have fewer gates and parameters than a brickwork circuit product spectrum ansatz.

\subsection{Complementary work}

Several prior works have looked at the relationship between MERA tensor networks and thermal states but taking rather different approaches than we take here. Examples include \cite{Czech_2016} where a quotient of a ground state MERA is taken to represent thermal states, or in \cite{Lin_2021} where MERA-like networks are optimized to disentangle a copy of the thermofield double state. 

As a simple circuit ansatz, the TMERA can be used in conjunction with other methods for state preparation.
If the Hamiltonian of interest has a gauge symmetry, then the TMERA could be fruitfully combined with \cite{davoudi_toward_2022}.
It could also be used as a component of methods like \cite{Gibbs_prep_2022,schuckert_probing_2022}, where an initial approximate Gibbs state is refined using methods derived from fluctuation theorems, with better performance for initial states with close to optimal free energy.

\acknowledgements
  CDW gratefully acknowledges the U.S. Department of Energy (DOE), Office of Science, Office of Advanced Scientific Computing Research (ASCR) Quantum Computing Application Teams program, for support under fieldwork proposal number ERKJ347. TJS gratefully acknowledges support by the DOE under Award Number DE-SC0019139, and useful discussions with Ning Bao and Stephen Jordan. The work of BGS is supported in part by the AFOSR under grant number FA9550-19-1-0360. 

\bibliographystyle{unsrt}
\bibliography{refs.bib}

\appendix

\section{Free Fermions}\label{app:free-fermion}

A Gaussian fermion state $\rho$ can be fully characterized by the covariance matrix
\begin{align}
  \label{eq:covmat}
  \begin{split}
    \Gamma_{j,k} = \frac{i}{2} \mathrm{tr}( \rho [\gamma_j, \gamma_k])
    \end{split}
\end{align}

For $n$ local fermionic modes $a_i$ we can define $2n$ Majorana modes 

\begin{align}
\label{eq:maj}
\gamma_{2i-1} &= \frac{a_i + a^\dagger_i}{2} \\
\gamma_{2i} &=  \frac{a_i - a^\dagger_i}{2} 
\end{align} 

The Majorana modes are Hermitian and obey the usual fermionic anticommutation relations ${\gamma_i,\gamma_j} = 2\mathbb{I}$, but  also $\gamma^2_i = \frac{I}{2}$.

A Gaussian fermion state satisfies $\Gamma^2 \leq -I$, with equality holding for pure states. The expectation value of quadratic fermion observables can be directly read from the covariance matrix, while the Wick's theorem allows higher order observables to be calculated in terms of the quadratic expectation values. 

Subsystems of a Gaussian state are defined by restricting the covariance matrix to a subset of the Majorana operators, eliminating the rows and columns associated with the discarded operators. Any mixed Gaussian state can be considered a subsystem of some pure Gaussian state. The covariance matrix for a mixed state can be purified by adding additional Majorana modes, introducing the new rows and columns with entries so that the new covariance matrix satisfies $\Gamma^2 = -I$.

Although we can relate the fermionic system to a system of qubits using a Jordan-Wigner duality, this operation of taking fermionic subsystems is a different restriction of the Hilbert space from the tracing out of qubit degrees of freedom. However, all even parity operators on the reduced fermionic and qubit subsystems will agree. Odd parity operators have zero expectation value in Guassian states. 

A mixed Gaussian state may also be thought of as a thermal Gibbs state for some quadratic Hamiltonian.

To write the Gibbs state of a general quadratic Majorana Hamiltonian, first write the Hamiltonian
\begin{equation}
    H = \sum_{ij} h_{ij} \gamma_i \gamma_j
\end{equation}
in terms of a real antisymmetric matrix of coupling constants $h$. 
The covariance matrix for the Gibbs state of this Hamiltonian at inverse temperature $\beta$ is then
 \begin{align}
  \label{eq:covtherm}
  \begin{split}
    \Gamma_\beta = i \tanh (i \beta h)
    \end{split}
\end{align}

For some mixed covariance matrix $\Gamma$ with eigenvalues $|\lambda_i | < 1$ then the Hamiltonian for which it is a Gibbs state is then given by $\tilde{h} = -i/\beta \mathrm{atanh}(-i \Gamma)$. 

A Guassian unitary transformation always exists which transforms a Gaussian state into a collection of independent fermion modes, each with excitation probability $p_j$, so every Gaussian state has a product spectrum.
 Gaussian unitary operations are equivalent to linear transformations on the fermion modes, which act on the covariance matrix via an orthogonal transformation $\Gamma' = O \Gamma O^T$. 
The Gaussian unitary transformation that decouples the Gaussian fermion state block diagonalizes the covariance matrix into two-by-two blocks with off diagonal entries $\pm i \lambda_j$. 

Local matchgate circuits are equivalent to unitary Guassian operations and can be implemented using covariance matrix methods. Local two-qubit gates become orthogonal transformations on four consecutive Majorana operators.

Because all Gaussian states have a product spectrum, the entropy of any mixed Gaussian state is the sum of entropies $S_j$ from the states eigenmodes, each of which can be written in terms of the respective eigenvalue of the covariance matrix. 

 \begin{align}
  \label{eq:covEnt}
  \begin{split}
   S_j &=  -  \frac{1 + \lambda_j }{2} \log \frac{1 + \lambda_j }{2} -  \frac{1 - \lambda_j }{2} \log \frac{1 - \lambda_j }{2} \\
   &= \log 2 - \log \sqrt{1 - \lambda_j^2} - \lambda_j \log \sqrt{\frac{1 + \lambda_j}{1 - \lambda_j}}
    \end{split}
\end{align}

We can also compute the fidelity between two mixed Gaussian states can also be computed in terms of their covariance matrices \cite{Swingle_2019},
 \begin{align}
  \label{eq:covFid}
  \begin{split}
   \mathcal{F}_{\rho, \sigma} =  2^{-\frac{n}{2}} \mathrm{det}(I - \Gamma_\rho \Gamma_\sigma)^{\frac{1}{4}} \mathrm{det}\left ( I - \sqrt{ I + \Gamma_{\rho,\sigma}^2} \right )^{\frac{1}{4}}
    \end{split}
\end{align}
with $\Gamma_{\rho,\sigma} = \frac{\Gamma_\rho + \Gamma_\sigma}{I - \Gamma_\rho \Gamma_\sigma}$. 

If at least one of these states is pure then the simpler expression can be used, $ \mathcal{F}_{\rho, \sigma} =  |\det( (\Gamma_\rho + \Gamma_\sigma)/2)|^{1/4}$.

The quadratic fermionic Hamiltonian we work with which is dual to the critical transverse-field Ising model is shown in Eq. \eqref{eq:f-ham}.

For finite sized spin chain with periodic boundary conditions, the dual quadratic fermion Hamiltonian must have anti-periodic boundary conditions for the even parity states, and periodic boundary conditions for odd parity states. Thus, with periodic boundary conditions the full Ising Hamiltonian is not strictly dual to a single quadratic fermion Hamiltonian, but all of its eigenstates are dual to a Gaussian fermion states.

 To avoid these issues we will focus just on the Majorana fermion model with anti-periodic boundary conditions for our numerics, and in this case our circuits may be interpreted as local operations on fermionic degrees of freedom. In fact, because we are working with a free-fermion model, we will also restrict our gates to be unitary Gaussian fermion operations, equivalent to Bogliubov transformations.
  In terms of spins these operations are dual to matchgate circuits, and can be simulated classically via Gaussian fermion methods. Although we will restrict our attention to noninteracting fermions, the methods can be straightforwardly applied to qubits on a quantum computer for preparing states of spin systems, and the local gates generalized beyond matchgates to target interacting models.

\section{Product spectrum ansatz as mean-field solution of a classical stat mech model}\label{app:psa-mean-field}

Our TMERA is a special case of the product spectrum ansatz.
In a product spectrum ansatz, one fixes a unitary $U$ and writes a state
\begin{align} \label{eq:psa-fermion}
    \rho = U \left[\bigotimes_j [p_{j1} n_j + p_{j0} (1-n_j)\right] U^\dagger
\end{align}
where $n_j$ is the fermion number operator on mode $j$.
Each mode has an independent occupation probability $p_{j1}$;
normalization requires $p_{j0} = 1 - p_{j1}$. 
In Sec.~\ref{s:ansatz-intuition} we wrote the density matrix \eqref{eq:psa-fermion}
in terms of Pauli matrices $\sigma^z_j = 2n_j-I$
and framed the occupation probability $p_{j1}$ in terms of the mode energy $E_j$.
In this context it is easier to think in terms of occupations and occupation probabilities;
we will see the mode energy emerge.
The density matrix $\rho$ of \eqref{eq:psa-fermion} can be written concisely in an occupation number basis $\ket{\bm n} = \ket{n_1,\dots, n_L}$ as
\begin{align} \label{eq:psa-fermion-fock}
    \rho = \sum_{\bm n} p_{\bm n} \ketbra{\bm n}{\bm n}\;, \qquad p_{\bm n} = \prod p_{jn_j}\;.
\end{align}
One then variationally minimizes the free energy $F = \expct{E} - TS$ over the probabilities $p_{jn}$. 

While variational minimization of the free energy is appropriate on a quantum computer,
it is easier to think about maximizing the entropy at fixed energy,
which comes in with a lagrange multiplier.
The entropy is straightforward: it is the sum of mode entropies
\begin{align}
 S = \sum_j S_j\;,\qquad S_j = \sum_n p_{jn} \ln p_{jn}\;.
\end{align}
This follows from the fact that $\rho$ is a product state in the mode basis,
or from explicit calculation on \eqref{eq:psa-fermion-fock}.

The energy expectation value $\expct{E}$ is somewhat more involved; 
it leads to an effective classical stat mech model on the mode occupation. Because density matrices of the form \eqref{eq:psa-fermion-fock} are diagonal in the mode occupation number basis $\ket{\bm n}$
we can write the energy expectation value
\begin{align}
    \expct{E} = \tr \rho H = \sum_{\bm n} E_{\bm n} p_{\bm n}
\end{align}
with energies
\begin{align}
    E_{\bm n} = \braket{\bm n | H | \bm n}\;.
\end{align}
We can compute this by expanding $H$ in mode creation and annihilation operators:
\begin{align} \begin{split}
    H =\quad &\sum_{jk} \Big[(T_{jk} c^\dagger_j c_k + T'_{jk} c^\dagger_j c^\dagger_k) + h.c.\Big] \\
    +&\sum_{jj'kk'} V_{jkj'k'} c^\dagger_jc^\dagger_k c_{j'} c_{k'} \\
    +&\cdots\;.
\end{split} \end{align}
Off-diagonal terms disappear, so
\begin{align}\label{eq:fock-energy}
    E_{\bm n} = \sum_j T_j n_j + \sum_{j<k} U_{jk} n_j n_k + \dots
\end{align}
where
\begin{subequations} \begin{align}
    T_j &= T_{jj}\\
    U_{jk}  &= V_{jkkj} - V_{jkjk}\;.
\end{align} \end{subequations}
Our problem is then to maximize the entropy of
the classical probability distribution $p_{\bm n}$ on occupation number states $\bm n = (n_1, \dots, n_L)$,
subject to a constraint $E = \sum E_{\bm n} p_{\bm n}$.
This is precisely a classical stat mech model.

When we impose the product spectrum ansatz \eqref{eq:psa-fermion-fock}, 
we disallow correlations between the modes.
Performing the entropy maximization within this ansatz leads to a self-consistent mean-field solution of the stat mech model given by the energies \eqref{eq:fock-energy}.
To see this, write the entropy maximization condition
\begin{align}
\begin{split}
    0 = \frac{\partial}{\partial p_{jn} }
    \Bigg[ &\sum_j \sum_{n_j} p_{jn_j} \ln p_{jn_j}   \\
    &+ \sum_j F_j \Bigg(1 - \sum_{n_j} p_{jn_j} \Bigg)     \\
    &+ \beta \Bigg(E - \sum_{\bm n} E_{\bm n} \prod_j p_{j n_j}\Bigg) \;.
\end{split}
\end{align}
The derivative $\frac{\partial}{\partial p_{jn}}$ is with respect to the probability that site $j$ has occupation number $n$;
the $L$ Lagrange multipliers $F_j$ ensure normalization of the mode probabilities $p_{jn}$
and the Lagrange multiplier $\beta$ ensures that the probabilities give the correct energy expectation value $E$.
Taking the derivative and performing some sums gives
\begin{align}
    0 = - \ln p_{jn} - F_j - \beta \Bigg(T_j n + \sum_l U_{jl} n_j \expct{n_l}\Bigg)\;,
\end{align}
together with an $n$-independent term we can sweep into $F_j$.
Here $\expct{n_l} = \sum_n n p_{ln}$ is the expectation value of the occupation number on site $l$.
The term in parentheses is exactly the mean-field energy
\begin{align}\label{eq:mf-energy}
    E_{\MF;jn} = T_j n + \sum_l U_{jl} n_j \expct{n_l}\;.
\end{align}
We can now write
\begin{align}
    p_{jn} = e^{-F_j - \beta E_{\MF;jn}}\;.
\end{align}
Since the $E_{\MF;jn}$ depend on the other mode probabilities $p_{ln}$,
we must solve this set of equations self-consistently---if we wish to do so directly.
By variationally optimizing the free energy, we short-circuit the self-consistent mean-field problem.

\section{Symmetrized State}
Applying random symmetries to the state can increase the entropy of a state which is not invariant under these symmetries, however, by definition they cannot change its energy. Applying this procedure to an approximate thermal state could only stand to increase the fidelity of that state, because the thermal state at a given energy is the maximal entropy state at that energy.

\section{Renormalization of temperature}
In preparing a state with a DMERA circuit, an intermediate state of smaller system size is prepared after each scale transformation, which can be thought of as a course grained version of the final state. For ground state preparation of a critical these intermediate states have nearly identical local expectation values and subsystem entropies to the final state due to the scale invariance.
However, for a Gibbs state the finite temperature breaks scale invariance, as we see from the different excitation probabilities at each scale. 
This is also reflected in the different energy density and correlation lengths of the intermediate states prepared during the course of the approximate Gibbs state preparation. These intermediate states may also be thought of as course grained versions of the final Gibbs state. 
Are these intermediate states also well described as thermal states at a higher temperature and smaller system size? Since we optimize the probabilities for a particular final system size its not necessary that the intermediate states would also be optimal. However, due to the scaling of the mode energies at different scales according roughly to $E_\ell \sim 2^\ell$, it's plausible that the state before the last scale transformation may look like a thermal state at twice the temperature.

\section{Hamiltonian of Approximate Gibbs State}
We can find a Hamiltonian for which our approximate Gibbs state is an exact thermal state by taking the logarithm of our density matrix, which is unambiguous up to an overall shift in the energies, so we can fix the trace of the Hamiltonian to be zero.
 For a high temperature state, the Hamiltonian is the leading order perturbation to the normalized identity matrix.
\begin{align}
  \begin{split}
    \rho_\beta = \frac{e^{-\beta H}}{\mathcal{Z_\beta}} \approx \frac{1}{L} ( I - \beta H)
  \end{split}
\end{align}
In terms of our covariance matrix formalism, an analogous procedure can be used to get a quadratic fermion Hamiltonian from one of our approximate thermal states. 
The coupling matrix of this Hamiltonian proportional to a high temperature thermal state approximation, so the features of its two-point correlation functions become features of the Hamiltonian for which it is a thermal state. 
This means the couplings of this Hamiltonian will be nonlocal put decaying roughly inverse polynomial with distance. 
The couplings at a fixed distance will also not be strictly equal due to broken translation invariance but instead vary spatially around mean values. 
\begin{align}
  \begin{split}
    H_\beta = \frac{-i}{\beta} \mathrm{atanh} (-i \tilde{\Gamma}_\beta)
  \end{split}
\end{align}
The spectrum of the modes is flat due to the degeneracy of excitation probabilities for all modes at a given scale. The wavepacket modes are themselves the eigenmodes of this Hamiltonian, and so they have compact spatial support. 
In the Heisenberg picture, the ground state DMERA circuit can be seen as transforming an initial Hamiltonian of only single-body terms to this long range Hamiltonian.
Time evolution according to this Hamiltonian is therefore equivalent to time evolution of the trivial decoupled Hamiltonian conjugated by the DMERA circuit.
This evolution will quickly deviate from evolution by the true local translational invariant Hamiltonian because the eigenmodes being evolved by the long range Hamiltonian are the wavepacketmodes which have compact support over varying lengthscales, and have a flat spectrum compared to the true Hamiltonian.

We also derive this Hamiltonian from a particular thermal state approximation, so in principle the approximate Gibbs states at different temperatures could give rise to different Hamiltonians. In principle it may be possible that for any given induced Hamiltonian, only its Gibbs state at the particular temperature of the Gibbs state approximation is a good analogue to the true thermal state of our target Hamiltonian. 
However, if these Hamiltonians induced at different temperatures do not deviate much, then choosing any one of them should give good approximations to our target thermal states for the whole range of thermal states. 

By the construction of the Hamiltonian, it is equivalent to a product state Hamiltonian conjugated by a logarithmic depth circuit. This mean time evolution by this Hamiltonian can be efficiently fast-forwarded \cite{Gu_2021}. This is true even in the case where the circuit is not a matchgate circuit and therefore is not equivalent to some quadratic fermion model.

\section{TFD purification}

A mixed state can be purified by adding additional degrees of freedom that the state is entangled with, such that the reduced state on the original subsystem is the same mixed state. with the additional Hilbert space dimension equal at least to the rank of the mixed state. The purified state may be written as a Schmidt decomposition in the energy eigenbasis with the coefficients equal to the square root of the normalized Boltzman weights. A natural choice of purifying system is a second copy of the same thermal state, since it has the minimal Hilbert space size and has the same eigenvalue spectrum. 

$$\psi_\beta = \frac{1}{\sqrt{\mathcal{Z_\beta}}} \sum_i e^{-\beta E_i /2} \ket{E_i}_L\ket{\tilde{E_i}}_R$$

We can prepare an ansatz state for the TFD using an extension of our TMERA circuit, where each mixed qubit is half of an entangled pair of qubits. Pairs of qubits destined for the two halves of the TFD state will be entangled to the degree required that their entanglement entropy matches the desired entropy for qubits at a given scale $\ell$ for the target temperature $T$. Two copies of the DMERA circuit are then applied to each collection of appropriately entangled qubits to produce an ansatz state for the TFD at temperature $T$. By the product-state nature of the TMERA ansatz, these TFD states will only have Bell-pair like entanglement between the two halves.

\end{document}